\documentclass[preprint,superscriptaddress,groupedaddress]{revtex4}
\usepackage{amsfonts,amssymb,amsmath,bm}
\usepackage{amscd}
\usepackage{amssymb}
\usepackage{graphicx}
\usepackage{theorem}
\usepackage{makeidx}
\usepackage{amsmath}
\usepackage{epsfig}
\usepackage{verbatim}
\usepackage{bm}

\usepackage{color}

\DeclareMathOperator{\sign}{sign}

\newcommand{\la}{\label}
\newcommand{\bbm}{\begin{multline}}
\newcommand{\eem}{\end{multline}}
\newcommand{\be}{\begin{equation}}
\newcommand{\ee}{\end{equation}}
\newcommand{\bea}{\begin{eqnarray}}
\newcommand{\eea}{\end{eqnarray}}
\newcommand{\p}{\partial}

\newcommand{\xv}{\bm{{\rm x}}}

\newcommand{\uv}{\bm{{\rm u}}}
\newcommand{\vv}{\bm{{\rm v}}}

\newcommand{\rv}{\bm{{\rm r}}}

\newcommand{\sech}{{\mathrm{sech}}}

\begin{document}
\title{Odd viscosity in chiral active fluids}

\date{\today}

\author{Debarghya Banerjee}
\affiliation{Instituut-Lorentz, Universiteit Leiden, Leiden 2300 RA, The Netherlands}

\author{Anton Souslov}
\affiliation{Instituut-Lorentz, Universiteit Leiden, Leiden 2300 RA, The Netherlands}

\author{Alexander G.~Abanov}
\affiliation{Department of Physics and Astronomy and Simons Center for Geometry and Physics, Stony Brook University, Stony Brook, NY 11794, USA}

\author{Vincenzo Vitelli}
\affiliation{Instituut-Lorentz, Universiteit Leiden, Leiden 2300 RA, The Netherlands}

\begin{abstract}
Chiral active fluids are materials composed of self-spinning rotors that continuously inject energy and angular momentum at the microscale.
Out-of-equilibrium fluids with active-rotor constituents have been experimentally realized using nanoscale biomolecular motors~\cite{Sumino2012, Tabe2003, Oswald2015,Drescher2009, Petroff2015, Riedel2005, Denk2016}, microscale active colloids~\cite{Snezhko2016, Maggi2015, Lemaire2008}, or macroscale driven chiral grains~\cite{Tsai2005}.
Here, we show how such chiral active fluids
break both parity and time-reversal symmetries in their steady states, giving rise to a dissipationless linear-response
coefficient called odd viscosity~\cite{Avron1995, Avron1998} in their constitutive relations.
Odd viscosity couples pressure and vorticity leading, for example, to density modulations within a vortex profile.
Moreover, chiral active fluids flow in the direction transverse to applied compression as in shock propagation experiments.
We envision that this collective transverse response may be exploited to design self-assembled hydraulic cranks that convert between
linear and rotational motion in microscopic machines powered by active-rotors fluids.
\end{abstract}

\maketitle

\begin{figure}[th!]
\includegraphics[angle=0]{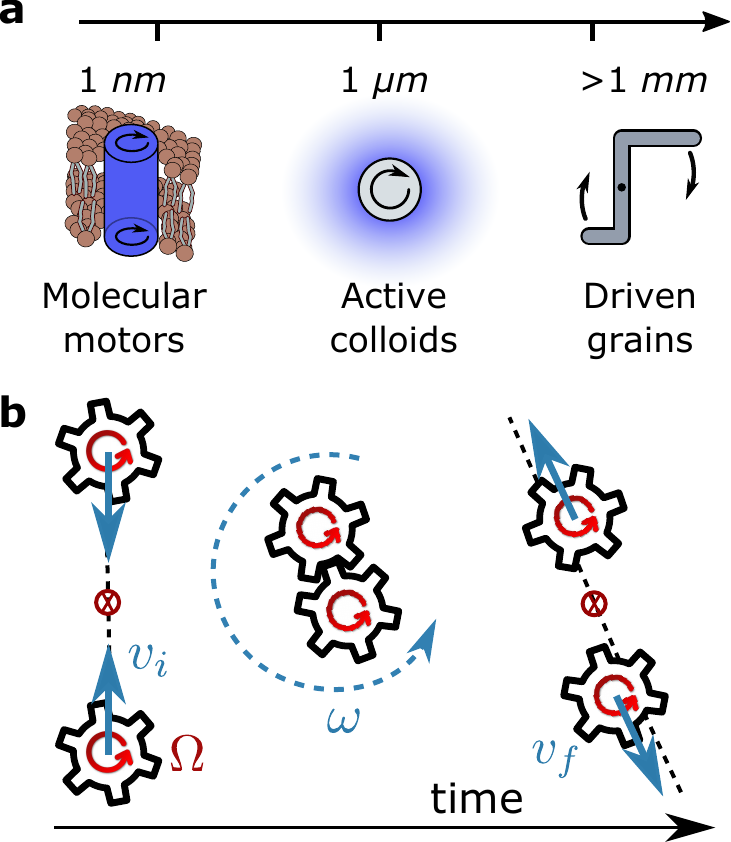}
  \caption{{\bf Chiral active liquids with odd viscosity.} 
(a) Chiral active liquids in a variety of contexts: biological~\cite{Sumino2012, Tabe2003, Oswald2015, Drescher2009, Petroff2015, Riedel2005}, colloidal~\cite{Snezhko2016, Maggi2015, Lemaire2008}, and granular~\cite{Tsai2005}. 
(b) Schematic of the collision processes in a chiral active gas. (Left) Head-on collision between self-spinning gears that initially move with speed $v_i$ and rotate with frequency $\Omega$. Their center of mass is represented as a red crossed circle. (Middle) While in contact, the frictional gears convert intrinsic angular momentum into orbital angular momentum, which leads to rotation around their center of mass with frequency $\omega$ (shown as a dashed circular arrow). We assume that this process occurs on a time scale which is fast compared to the time between collisions. As a result, the spinning frequency is rapidly reset to the initial $\Omega_0$ favored by the balance of internal active torque and dissipation. (Right) After the collision, the self-spinning gears move away from each other with velocity $v_f$. However, the particles do not necessarily move with final velocities that parallel the vector distance between them, i.e., orbital angular momentum is generated in the collision.
}
\label{Fig1}
\end{figure}

The mechanical response of any viscoelastic material is encoded in its constitutive relations: a set of equations
that express the stress tensor in terms of the strain and strain rate~\cite{LandauVI}. Conservation of angular momentum dictates that 
the stress tensor $\sigma_{ij}$ of any medium with vanishing bulk external torque
must be symmetric under the exchange of its two indices $i$ and $j$. 
This conclusion, however, does not apply to chiral fluids composed of self-spinning constituents (see Fig. \ref{Fig1}a) that are driven by active torques~\cite{Lenz2003, Uchida2010, Yeo2015, Spellings2015, Nguyen2014, vanZuiden2016}. 
In addition to the presence of an {\it antisymmetric} stress~\cite{Dahler1961, vanZuiden2016, Condiff1964, Tsai2005,Bonthuis2009,Furthauer2012}, 
chiral active media exhibit anomalies in the {\it symmetric} component of $\sigma_{ij}$ that encodes the viscous stress.

In this Letter, we ask a deceptively simple question: What is the viscosity of a chiral active fluid?
Viscosity typically measures the resistance of a fluid to velocity gradients. It is expressed mathematically by a tensor, $\eta_{ijkl}$, that acts as a coefficient of proportionality between viscous stress
and strain rate~\cite{LandauVI}. The Onsager reciprocity relation stipulates that $\eta_{ijkl}$, like any linear transport coefficient, must be symmetric (or even) under the exchange of the first and last pairs of indices (i.e., $\eta_{ijkl}=\eta_{klij}$)
provided that time-reversal symmetry holds~\cite{LandauVI}.
Here, we show that chiral active fluids acquire an additional odd (or Hall) viscosity $\eta^o_{ijkl}(=-\eta^o_{klij})$, as a result
of the breaking of both parity and time-reversal symmetries.

\begin{figure}[th!]
\includegraphics[angle=0]{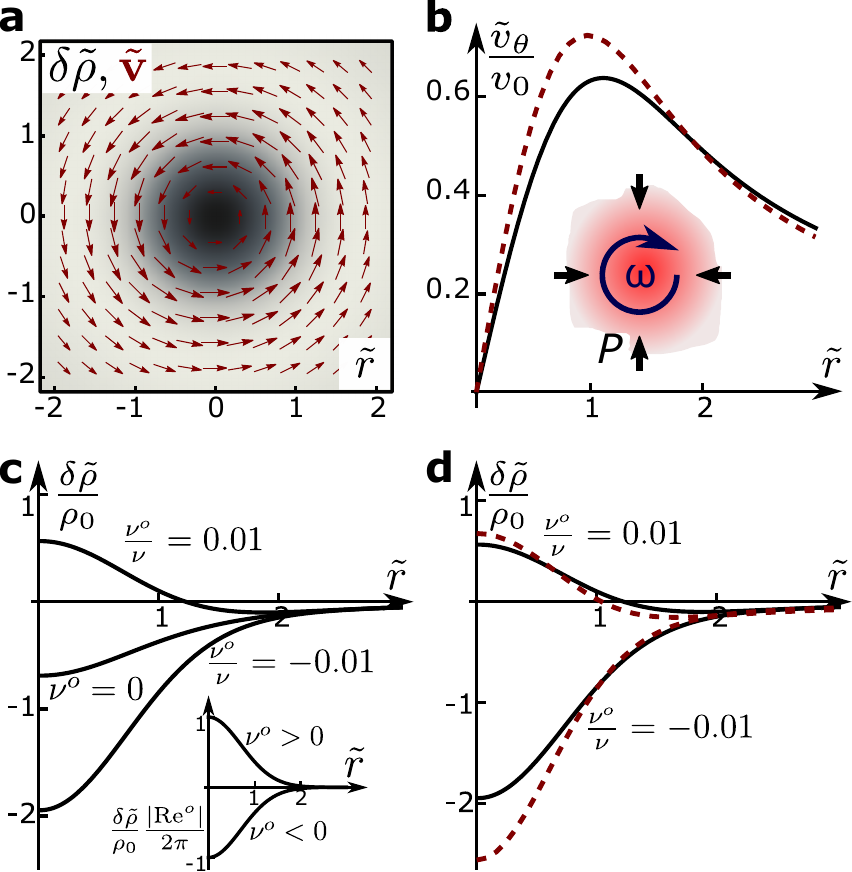}
  \caption{{\bf Odd viscosity in weakly compressible active fluids can lead to a
build-up of particles inside a vortex.} 
The Lamb-Oseen vortex flow (a: red
arrow, b: Rescaled azimuthal component $\tilde{v}_\theta \equiv v_\theta r_s \pi/r_0$, $r_s \equiv r_{0}\sqrt{1 + t/t_{0}}$, and $\tilde{r} \equiv r/r_s$) is unperturbed by the presence of odd viscosity $\nu^o$ (b: black, solid), but does change due to antisymmetric stress (b: red, dashed). 
[b inset: schematic of coupling between density and vorticity in odd-viscosity fluids.]
Odd viscosity does change the density profile (grayscale map of a; c-d). For $\frac{\nu^o}{\nu}=0.01$, excess density builds up at the vortex core and can dominate over the usual density depletion in an \emph{inertial} vortex when $\mathrm{Re}^o \equiv v_0 r_0/\nu^o \ll 1$. We plot time-rescaled solutions for odd-viscosity vortices of particle density $\delta \tilde{\rho}/\rho_0 \equiv (\rho - \rho_0) \frac{\pi^2 (1+t/t_0)}{\rho_0 \mathrm{Ma}^2} $, as a function of radius $\tilde{r} \equiv r/r_s$ for Reynolds number $\frac{v_0 r_0}{\nu} = 0.05$ and $\frac{\nu^o}{\nu} = \{-0.01, 0, 0.01\}$, (see SI for details). [Inset: density plot in the limit $Re^o \ll 1$, see Eq.~(\ref{eq:r})]. (d) Snapshot of density profile without (black, solid) and with (numerical, red, dashed) effects of a moderately small antisymmetric stress.}
\label{Fig2}
\end{figure}

Avron et al.~first recognized that a two-dimensional electron fluid can display a Hall viscosity in the presence of an external magnetic field that breaks time-reversal symmetry (TRS) at 
equilibrium~\cite{Avron1995,Avron1998,Read2009,Wiegmann2014,Lapa2014,Moroz2015}. In chiral active fluids, violation of Onsager reciprocity originates from the breaking of microscopic reversibility out of equilibrium, 
a feature inherent to active matter~\cite{Marchetti2013, Solon2015}. In this case,
an odd viscosity can emerge as a linear-response coefficient calculated around the non-equilibrium steady state of a purely classical system. 
Despite its universal nature, odd viscosity was neglected in previous hydrodynamic theories of active rotors~\cite{Dahler1961, vanZuiden2016, Condiff1964, Tsai2005,Bonthuis2009,Furthauer2012} that consider only rotors with small spinning frequency---a regime for which the antisymmetric stress dominates over the odd viscosity.
On general grounds~\cite{Read2009, Wiegmann2014}, it can be shown that odd viscosity is proportional to the non-vanishing angular momentum density which exists within the active fluid in steady state.

For concreteness, we construct a hydrodynamic description of dry chiral active fluids based on a constitutive relation that explicitly accounts for an odd viscosity term. In two dimensions, the evolution of the three slow variables, i.e., density of particles $\rho$, linear momentum $g_i \equiv \rho v_i$, and intrinsic angular momentum $\ell$, is governed by the following equations (see SI for detailed derivations):
\begin{align}
	 D_t \rho &= 0\,,
 \label{eq-rhou2} \\
	 D_t \ell &= \tau+D^\Omega \nabla^2\Omega
	- \Gamma^\Omega \Omega
	-\epsilon_{ij}\sigma_{ij} \,,
 \label{eq-lu2} \\
	D_t g_i &= \partial_{j}\sigma_{ij}-\Gamma^v v_i \,.
 \label{eq-rhouu2} 
\end{align}
where $D_t$($= \partial_t + v_k \partial_k$) denotes a convective derivative and $\ell \equiv I \Omega$ can be expressed in terms of the local spinning frequency of the rotors $\Omega(\xv,t)$ and
the moment-of-inertia density $I$. A constant active torque applied to each rotor and described by the torque density $\tau$ injects energy into the fluid, thus breaking detailed balance.
The rotational dissipation coefficient $\Gamma^{\Omega}$ saturates energy injection in a way that leads to a non-equilibrium steady state with characteristic single rotor frequency $\Omega \sim \tau/\Gamma^{\Omega}(\equiv \Omega_0)$, whereas $D^\Omega$ and $\Gamma^{v}$ control diffusion of intrinsic rotations and linear momentum damping, respectively. While the damping $\Gamma^v$ might be significant in many realizations of active rotor systems, we omit it for simplicity in this work.

The stress in Eqs.~(\ref{eq-lu2}-\ref{eq-rhouu2}) is given by
\begin{equation} 
 \label{eq:ss}
	\sigma_{ij} \equiv \epsilon_{ij} \frac{\Gamma}{2} 
	\left(\Omega  - \omega\right)  - p \delta_{ij} + \eta_{ijkl} v_{kl} 
	+ \frac{\ell}{2}(\partial_i v_j^*+\partial_i^* v_j),
\end{equation} 
where $\omega \equiv \frac{1}{2} \epsilon_{ij} \partial_i v_j$ is the vorticity
and $v_j^*\equiv \epsilon_{jl} v_l$ is the velocity vector rotated clockwise by $\pi/2$. (Note that $\epsilon_{ij}$ denotes the Levi-Civita antisymmetric tensor in 2D.)
The antisymmetric term in Eq.~(\ref{eq:ss}) proportional to $\Gamma$ results from inter-rotor friction and couples the flow $\vv$ to the intrinsic rotations $\Omega$~\cite{Tsai2005}. 
The last component of the stress $\sigma_{ij}$  is the novel ingredient that we add to Eq.~(\ref{eq:ss}), see SI for a variational hydrodynamics derivation. This term is a nonlinear coupling between the fields $\vv(\xv,t)$ and $\ell(\xv,t)$ that was neglected in previous hydrodynamic theories~\cite{Dahler1961, Condiff1964, Tsai2005,Bonthuis2009,Furthauer2012, vanZuiden2016}.
If $\ell$ is fixed (e.g., a constant), this nonlinear coupling reduces to a transport coefficient $\eta^o = \ell/2$ called odd viscosity
in the constitutive relations, Eq.~(\ref{eq:ss}).

For example, if the active torque dominates over the inter-rotor coupling $\Gamma$, the ensemble of self-spinning rotors behaves as a weakly interacting chiral active gas. In such a gas, the active rotation frequency is near $\Omega_0$ for each particle except during and immediately after each collision, when some intrinsic rotation is converted into fluid vorticity by the antisymmetric stress (see Fig. \ref{Fig1}b). 
While this conversion is crucial in establishing the chiral steady state of a gas of rotors, the state itself depends only on odd viscosity $\eta^o$ and not on the inter-rotor coupling, if $\Gamma$ is sufficiently small.

We now determine the conditions for the emergence of odd viscosity.
Gradients of intrinsic angular rotation $\Omega$ are negligibly small if the characteristic velocity $v_0$ and length scale $r_0$ are such that $\Gamma/I \ll v_0/r_0 \ll \tau/ \Gamma$. 
For this to hold, a necessary condition is $\tau \gg \Gamma^2/ I$. In this regime, Eq.~(\ref{eq-rhouu2}) decouples from Eq.~(\ref{eq-lu2}) and becomes the modified Navier-Stokes equation: 
\begin{equation}
D_t v_i = \nu \nabla^2 v_i +  \nu^{o}  \nabla^2 \epsilon_{ij} v_j - \frac{\partial_i p}{\rho},\label{eq:lin}
\end{equation}
with a familiar kinematic viscosity $\nu$($\equiv \eta/\rho$) and an additional odd viscosity $\nu^o$($\equiv \eta^{o}/\rho$) term.
The field $\Omega(\xv,t)$ has been integrated out from Eq. (\ref{eq:lin}): the only vestige of its presence is the emergent transport coefficient $\nu^o$. 
 Leading-order corrections to Eq.~(\ref{eq:lin}) in gradients of $\Omega$ are captured by the antisymmetric stress, the first term in Eq.~(\ref{eq:ss}). 
The effective theory embodied by Eq.~(\ref{eq:lin}) ceases to be valid whenever large spatial gradients of $\Omega(\xv,t)$ are created by interactions between rotors 
(e.g., at large densities) -- in that case we resort to the full Eqs.~(\ref{eq-rhou2}-\ref{eq-rhouu2}).

Inspection of Eq.~(\ref{eq:lin}) reveals that the odd viscosity term is a \emph{transverse} linear-response coefficient describing forces $f_i$ due to gradients 
in the perpendicular flow components $\epsilon_{ij} v_j$. 
In addition, $\nu^o$ is odd under either parity $P$ or time-reversal $T$ symmetries: $P \nu^o = T \nu^o = - \nu^o$ and, thus, it is nonzero only if both $P$ and $T$ are broken.
Thus, the odd viscosity term $\nu^o \nabla^2 \epsilon_{ij} v_j$ is $T$-invariant and thereby reactive: unlike dissipative viscosity $\nu$, odd viscosity $\nu^o$ is not associated with energy dissipation~\cite{Avron1998}.
Significantly, the derivation presented in the SI for the odd viscosity of a chiral active gas relies only on conservation laws: it does not require dissipation.

As in ordinary hydrodynamics, two familiar dimensionless parameters can be used to classify different phenomena described by Eq.~(\ref{eq:lin}): (i) the Reynolds number $\mathrm{Re} \equiv v_0 r_0/\nu$ and (ii) the Mach number $\mathrm{Ma} \equiv v_0/c$, where the speed of sound $c$ enters via $c \equiv \sqrt{\partial p/\partial \rho}$. 
For active fluids, speed of sound is associated with interactions between constituent rotors, and high Mach numbers may be achievable even in table-top active-fluid experiments. 
In the presence of odd viscosity, we need an additional dimensionless parameter: either the viscosity ratio $\nu^o/\nu$ or the \emph{odd} Reynolds number $\mathrm{Re}^o \equiv v_0 r_0/ \nu^o$.

\begin{figure}[t!]
\includegraphics[angle=0]{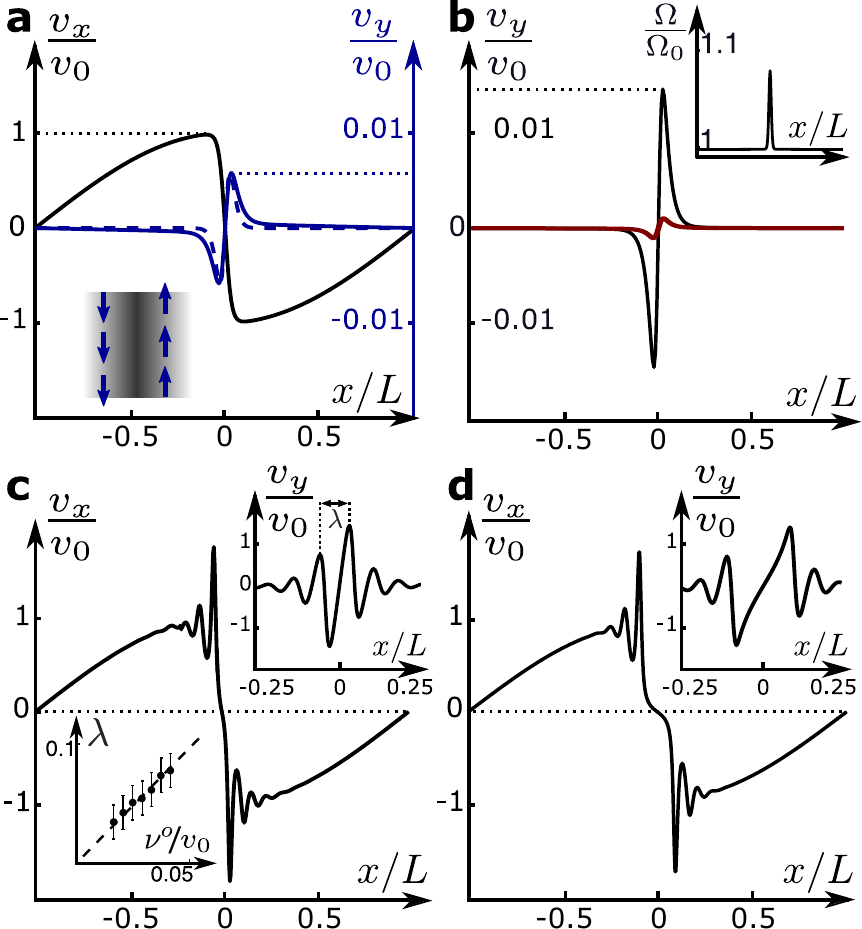}
  \caption{ {\bf Shocks in chiral active fluids.} (a) Inset: Shock generated by the external forcing ${\mathbf f}(\xv) = - \hat{x} \frac{f_0 \pi}{4} \sin \pi x/L$ exhibits transverse flow $v_y$ (blue arrows).
Main panel: longitudinal [black] and transverse [blue: analytic (dashed), numeric (solid)] flow profiles for small viscosity ratio $\nu^o/\nu = - 0.02$.
The characteristic velocity scale is $v_0 \equiv \sqrt{f_0 L/\rho_0}$; transverse flow $v_y$ scales as $v_0 \frac{\nu^o}{\nu}$.
(b) Even in the presence of antisymmetric stress,
the effects of odd viscosity can dominate if the variation in $\Omega$ is small: $|\Omega - \Omega_0|/\Omega_0 \lesssim 10\%$ (inset: $\Omega/\Omega_0$). 
The sharper peak (black) includes the combined effect of antisymmetric stress and odd viscosity.
Neglecting odd viscosity, we find only a small transverse flow due to antisymmetric stress (red).
Parts (c-d) show the longitudinal (main panels) and transverse (top insets) flow for $\nu^o/\nu = - 10$. 
Shocks develop oscillations with wavelength $\lambda \sim |\nu^o|/v_0$, see bottom inset of (c) for numerical verification of this scaling law.
In part (c), the antisymmetric stress is zero, whereas in (d) its value is as in part (b).
}
\label{Fig3}
\end{figure}

When the flow is incompressible, the odd viscosity can be absorbed into a rescaled pressure: $p \rightarrow p - 2 \eta^o \omega$~\cite{Avron1998}.
At low Mach and Reynolds numbers, we show that odd viscosity can replace inertia to stabilize a vortex, with a density peak at its core.
In Fig.~2a, we plot the Lamb-Oseen vortex flow profile~\cite{LandauVI} (which does
not depend on the value of odd viscosity), obtained from Eq.~(\ref{eq:lin}), and the profile of density variations at the center of the vortex.
In the presence of odd viscosity, density deviates from the Lamb-Oseen profile, see SI. The density variation $\delta \rho \equiv \rho(x,t) - \rho_0$, measured relative to its constant value $\rho_0$ away from the vortex core, is controlled by the odd Reynolds number and given by:
\begin{equation}
\label{eq:r}
\frac{\delta \rho}{\rho_0} = \frac{ 2 \mathrm{Ma}^2 }{\pi (1 + t/t_0) \mathrm{Re}^o}  e^{-\frac{r^2}{ 4 \nu (t_0 + t)}},
\end{equation}
where $t_0 \equiv r_0^2/(4 \nu)$.
In Fig.~2b, we show that for the case in which the rotational handedness of the vortex is aligned with the spinning direction of the rotors (i.e., $\mathrm{Re}^o > 0$),
the center of the vortex experiences an increase in pressure and a resulting excess particle density. Contrast this scenario with
the Lamb-Oseen solution ($\nu^o \rightarrow 0$), in which inertia causes a pressure dip and particle depletion at the vortex center. 
This effect, familiar from the physics of cyclones, is amplified when $\mathrm{Re}^o < 0$.
In Fig.~3c, we numerically compute vortex dynamics in the presence of antisymmetric stress [i.e., the full hydrodynamic Eqs.~(\ref{eq-lu2}-\ref{eq-rhouu2}), for details see SI], starting from an initial Lamb-Oseen profile. 
We find that although the addition of moderately small antisymmetric stress quantitatively corrects the flow and density profiles, 
the relative rotational handedness of the vortex still determines the characteristic density peak or trough at the vortex center. 

At high Mach number, we consider strong effects of compressibility
and obtain analytical and numerical solutions for modified Burgers' shocks in chiral active fluids. In ordinary fluids, where $\nu^o=0$, Burgers' shocks propagate along the direction of compression ($\hat{x}$ in Fig.~\ref{Fig3}) and are stabilized by the balance between dissipation and non-linearities in the convective derivative of Eq.~(\ref{eq:lin}).
We find that in the presence of odd viscosity, ultrasonic shocks contain an additional
flow transverse to the direction of shock propagation (Fig.~\ref{Fig1}c). 
Using the exact solution of the one-dimensional Burgers' equation, we find an analytical expression for transverse flow
in the regime of low viscosity ratio (see Fig.~\ref{Fig3}a and SI for derivation). The analytical profile of transverse flow (dashed blue line) agrees well with numerical solutions of chiral-active-fluid hydrodynamics (solid blue line). Note that the transverse flow (in particular, vorticity) is localized within the shock where $\nabla \cdot \vv$ is largest. The
familiar \emph{nonlinear} transport equation for the vorticity $\omega$ is modified as (see SI):
\begin{equation}
	\partial_t \omega + \nabla \cdot (\omega \vv) = \nu \nabla^2 \omega + \frac{\nu^o}{2 } \nabla^2 (\nabla \cdot \vv)\ ,
 \label{eq:numfig3-6}
\end{equation}
where the additional source term, proportional to $\nu^o$, vanishes if the fluid is incompressible, i.e., away from the shock where $\nabla \cdot \vv \rightarrow 0$. Note that this mechanism of generation of vorticity is independent of the inter-rotor friction captured by the antisymmetric stress which was explicitly neglected in writing Eq.~(\ref{eq:numfig3-6}).  
If a moderately small $\Gamma$ is introduced, the transverse flow acquires gradients in $\Omega$ (inset of Fig.~\ref{Fig3}b), but only small quantitative corrections to the flow profile, Fig.~\ref{Fig3}b.

More striking phenomena occur in the regime in which odd viscosity dominates over dissipation, i.e., $|\nu^o| \gg \nu$. In this case, the shock profile changes qualitatively (contrast Figs.~3a and~3c). The odd viscosity introduces strong dispersive effects that, in addition to dissipation, give rise to non-linear waves reminiscent of KdV-Burgers' shocks~\cite{whitham2011linear,Kulkarni2012}. Consistent with this interpretation, the numerically obtained profile for the transverse response in Fig.~3a can be characterized in terms of oscillations of wavelength $\lambda \sim |\nu^o|/v_0$ that decay over distance $\Lambda\sim|\nu^o|\lambda/\nu$ (see SI for derivation and bottom inset of Fig.~3c for numerical results). Figure~3d illustrates the small effect of the antisymmetric stress.

Two-dimensional incompressible inviscid fluids composed of many
interacting vortices have been previously proposed as examples of emergent odd viscosity fluids \cite{Wiegmann2014}. However, unlike particles with active rotations, vortices in real fluids are not stable
steady-state constituents, unless their circulation is quantized or viscosity
is zero (as in superfluids). Chiral active fluids are also significantly different from electron fluids because they exhibit an odd viscosity that (i) arises
only out of equilibrium, 
(ii) is always accompanied by an antisymmetric stress,
and (iii) is not well defined as particles jam and active rotations are hindered by interactions.
A chiral active fluid carries a crank mechanism within itself---it can convert between linear and rotational motion. We envision that this collective mechanical response could be exploited in self-assembled hydraulic devices and microscopic machines based on active components.

\textbf{Acknowledgments}: We thank D.~Bartolo, S.~Ganeshan, W.~Irvine, A.~Levine, G.~Monteiro, P. Wiegmann, and F.~Ye for stimulating discussions. A.S., D.B., and V.V. were funded by FOM, NWO (Vidi grant), and the Delta Institute for Theoretical Physics. A.G.A. acknowledges the financial support of the NSF under grant no. DMR-1606591 and the hospitality of the Kadanoff Center for Theoretical Physics.

A.S. and D.B. contributed equally to this work.

\pagebreak

{\Large Odd viscosity in chiral active fluids: Supplementary Information}
\vspace{2em}

\setcounter{equation}{0}
\setcounter{figure}{0}

\renewcommand{\theequation}{S\arabic{equation}}
\renewcommand{\figurename}{{\bf Fig. }}
\renewcommand{\thefigure}{{\bf S\arabic{figure}}}

\section{Variational principle for hydrodynamics with odd viscosity}
\label{sec:var}

The goal of this section is to introduce the variational principle for the hydrodynamics of a two-dimensional fluid characterized by the density of intrinsic angular momentum $\ell$. We assume that the internal energy density of the fluid $\epsilon(\rho,s,\ell)$ depends on the mass density $\rho$, entropy density $s$ and the density of intrinsic angular momentum $\ell$. Standard thermodynamic formulas give the following relations between $\epsilon(\rho,s,\ell)$, pressure $p$, chemical potential $\mu$, temperature $T$, particle mass $m$, and intrinsic angular velocity $\Omega$:
\bea
	p &=& \rho\epsilon_\rho+s \epsilon_s+\ell\epsilon_\ell-\epsilon \,,
 \\
 	\mu/m &=& \epsilon_\rho \,,
 \\
 	T &=& \epsilon_s \,,
 \\
 	\Omega &=& \epsilon_\ell \,.
\eea
Here $\epsilon_\rho=\p \epsilon/\p \rho$ etc.

Now we consider the fluid in local equilibrium characterized by space and time dependent fields $\rho(x,t)$, $s(x,t)$, $\ell(x,t)$ and fluid velocity $v_i(x,t)$ ($i=1,2$). We promote thermodynamics to hydrodynamics using the variational principle defined by the following action:
\bea
	S &=& -\int d^2x\,dt\, \Big[\xi_0
	+v^i\xi_i -  \frac{\rho v_i v^i}{2} +\epsilon(\rho,s,\ell)- \omega \ell\Big] \,,
 \la{action1}
\eea
where the fluid vorticity $\omega$ is defined via $\omega \equiv \frac{1}{2} \epsilon^{ij} \p_i v_j$. Note the unusual proportionality factor of $1/2$ in the definition of vorticity, which we use throughout the work. The vorticity $\omega$ defined in this way is the local angular velocity of the fluid rotation.
In (\ref{action1}), we used the notation:
\bea
	\xi_\mu \equiv \rho\p_\mu\theta +s\p_\mu\eta +\ell\p_\mu\phi +\Phi_\alpha \p_\mu \Psi_\alpha \label{cleb}
\eea
with $\mu=0,1,2$. The action (\ref{action1}) should be considered as a functional of the following independent fields $\rho, \theta, s, \eta, \ell, \phi, \Phi_\alpha, \Psi_\alpha$ and $v_i$. The parameters $\rho, \theta, \ldots$ are so-called Clebsch parameters~\cite{Zakharov1997}. We have added auxiliary pairs of Clebsch parameters $\Phi_\alpha, \Psi_\alpha$ with $\alpha=1,2,\ldots$ which are useful to describe generic hydrodynamic flows but will not play any role in the following.

The action (\ref{action1}) is standard except for the last term. This term explicitly breaks parity as it depends on $\epsilon^{ik}$. Below we will see that it leads to non-vanishing odd viscosity. The physical manifestations of this term are the main subjects of this paper.

Varying over fields $\theta, \eta, \phi$ we obtain the following conservation laws
\bea
	\p_t\rho+\p_i(\rho v^i) &=& 0\,,
 \la{eq-rhou} \\
	\p_t s+\p_i(s v^i) &=& 0\,, 
 \la{eq-su} \\
	\p_t \ell+\p_i(\ell v^i) &=& 0\,.
 \la{eq-lu} 
\eea

The velocity field $v_i$ in (\ref{action1}) is not a dynamical field. Varying (\ref{action1}) with respect to $v_i$ relates it to Clebsch parameters (\ref{cleb}):
\bea
	\rho v_i  &=& \xi_i -\frac{1}{2}\p_i^* \ell \,.
\eea
Taking variations of (\ref{action1}) with respect to all other fields, after some manipulations, one can arrive to the following equation of motion:
\bea
	\p_t (\rho v_i) + \p_j\Big[\rho v_i v_j +p\delta_{ij}
	-\sigma_{ij}^{odd}
	\Big] =0\,
 \la{eq-rhouu}
\eea
with
\bea
	\sigma_{ij}^{odd} = \eta^{o}(\p_i v_j^*+\p_i^* v_j),
 \la{sigmaodd1}
\eea
where
\bea
 	\eta^o = \frac{1}{2} \ell(\xv,t), \la{eq:igl}
\eea
which is the main result of this section. Note that $\eta^{o}$ in (\ref{sigmaodd1}) denotes the intrinsic angular momentum field $\ell(\xv,t)$, see Eq.~(\ref{eq:igl}). 
It reduces to an anomalous transport coefficient (referred to as odd viscosity) when the angular momentum density $\ell$ is fixed (e.g., a constant).  In general, interactions may generate higher order corrections to the variational functional, as discussed below and in Sec.~\ref{sec:hyd}.
However, the result in Eq.~(\ref{eq:igl}) holds exactly in the case of a weakly interacting chiral granular gas with fast local rotations. 
Here and in the following we use the notation $a_i^*\equiv \epsilon^{ik}a_k$. From (\ref{eq-rhouu}) we identify the quantity
\bea
	g_i \equiv \rho v_i 
\eea
as a momentum density of the fluid and $\sigma_{ij}^{odd}$ as a part of a viscous stress tensor of the fluid.

An important remark is in order. We notice that three equations (\ref{eq-rhou},\ref{eq-su},\ref{eq-lu}) and two components of equation (\ref{eq-rhouu}) give us five equations sufficient to determine five independent fields $\rho,s,\ell, v_i$. We define these equations as a complete system of hydrodynamic equations. The number of hydrodynamic fields (five) is smaller then the number of fields in the variational principle (\ref{action1}). This reduction is known as symplectic or Hamiltonian reduction.  

Another important remark is that the equations (\ref{eq-rhou},\ref{eq-su},\ref{eq-lu},\ref{eq-rhouu}) are necessarily dissipationless as they are derived from the time translational invariant action (\ref{action1}). Indeed, it is easy to derive from these equations the energy conservation law:
\bea
	\p_t\left[\epsilon +\frac{\rho v_k v^k}{2}\right]
	+\p_i\left[(\epsilon+p)v_i +\frac{\rho v_k v^k}{2}v_i
	-\sigma_{ij}^{odd}v_j\right] =0 \,.
 \la{en-cons}
\eea

\section{Dissipation and gradient corrections}
\label{sec:diss}

Let us consider the derivation of the energy conservation in more detail. 
We proceed as follows
\bea
	&& \p_t\left[\epsilon +\frac{\rho v_k v^k}{2}\right]
	+\p_i\left[(\epsilon+p)v_i +\frac{\rho v_k v^k}{2}v_i
	+\sigma_{ij}v_j\right] 
 \nonumber \\
 	&=& -\sigma_{ij}\p_i v_j 
	+\left(\epsilon_\rho - \frac{u_j^2}{2}\right)\Big[\p_t\rho+\p_i(\rho v_i)\Big]
	+\epsilon_s\Big[\p_t s +\p_i(s v_i)\Big]
	+\epsilon_\ell\Big[\p_t \ell +\p_i(\ell v_i)\Big]
 \nonumber \\
	&+& v_i\Big[\p_t (\rho v_i) +\p_j\Big(\rho v_i v_j +\delta_{ij}p-\sigma_{ij}\Big)\Big]\,.
 \la{energy-aux}
\eea
So far we did not use any equation of motion. Using (\ref{eq-rhou},\ref{eq-su},\ref{eq-lu},\ref{eq-rhouu}) in the right hand side and the form of odd viscosity tensor (\ref{sigmaodd1}) we  obtain the energy conservation ({\ref{en-cons}). Our goal is to generalize hydrodynamic equations in the presence of dissipation in such a way that the entropy production is explicitly positive and the energy is conserved up to the loss due to the work of external forces.

We assume the following modified equations of motion
\bea
	\p_t\rho+\p_i(\rho v_i+J_i^\rho) &=& 0\,,
 \la{eq-rhou1} \\
	\p_t \ell+\p_i(\ell v_i+J_i^\ell) &=& \tau
	-\epsilon_{ij}\sigma_{ij}- \Gamma^\Omega \Omega \,,
 \la{eq-lu1} \\
	\p_t (\rho v_i) + \p_j\Big[\rho v_i v_j +p\delta_{ij}
	-\sigma_{ij}
	\Big] &=& -\Gamma^v_{ij} v_j\,,
 \la{eq-rhouu1} \\
	\p_t s+\p_i(s v_i+J_i^s) &=& \frac{Q}{T}\,. 
 \la{eq-su1} 
\eea
Here, $\tau$ is the density of external torque acting on intrinsic rotational degrees of freedom of the fluid, $J_i^{\rho,s,\ell}$ are gradient corrections to currents, $\Gamma^{\Omega}$ is the intrinsic angular momentumdamping rate, and $Q$ is a heat production rate due to various friction forces. We consider the term $-\Gamma^v_{ij} v_j$ to be a linear-momentum damping term resulting from friction of the particles with a substrate, and therefore to have the form $\Gamma^v_{ij} = \Gamma^v \delta_{ij}$. In principle, $\Gamma^v_{ij}$ could also include a Lorentz-like component proportional to $\epsilon_{ij}$, for example in a high-Reynolds number flow as a result of a Magnus force. We do not consider the effects of these $\Gamma^v_{ij}$ components in this work.
The combination of (\ref{eq-lu1}) and (\ref{eq-rhouu1}) produces the conservation of angular momentum density
\bea
	\p_t(\epsilon^{ki}x_k \rho v_i +\ell) 
	+\p_j\Big(\epsilon^{ki}x_k\left[\rho v_i v_j +p\delta_{ij}
	-\sigma_{ij}\right]+\ell v_j+J_j^\ell\Big)
	= \tau -\Gamma^v \epsilon^{ki}x_k v_i -\Gamma^\Omega \Omega \,,
\eea
where the right hand side is the density of net torque acting on the system due to external forces.

Let us now fix all constitutive relations to the first order in gradients~\cite{Lucas2014} making sure that the heat production rate $Q$ is non-negative. We write
\bea
	J_i^\rho &=& -D^\mu \p_i\left(\mu - \frac{v_k^2}{2}\right) 
	+L^\mu \p_i^*\left(\mu - \frac{v_k^2}{2}\right) \,,
 \la{Jirho} \\
 	J_i^s &=& -D^T \p_i T +L^T \p_i^* T \,,
 \la{Jis} \\
 	J_i^\ell &=& -D^\Omega \p_i\Omega + L^\Omega \p_i^*\Omega\,,
 \la{Jil} \\
 	\sigma_{ij} &=& \sigma_{ij}^s +\frac{1}{2}\sigma^a \epsilon_{ij} \,,
 \la{sigmaij} \\
	\sigma^a &=& \Gamma \left(\Omega -\frac{1}{2}\p_k v_k^*\right)\,,
 \la{sigmaa}\\
 	\sigma_{ij}^s &=& 2\eta (\p_i v_j+\p_j v_i-\delta_{ij}\p_k v_k) + \delta_{ij}\eta^b (\p_k v_k)
	+ \eta^{odd}(\p_i v_j^*+\p_i^{*} v_j)
	\,,
 \la{sigmaijs} 
\eea
and for the total heat production rate
\bea
	Q &=& D^\mu \left[\p_i\left(\mu - \frac{v_k^2}{2}\right)\right]^2+D^T (\p_i T)^2
	+ D^\Omega (\p_i\Omega)^2
 \nonumber \\
	&+& \Gamma \left(\Omega -\frac{1}{2}\p_k v_k^*\right)^2
	+\eta (\p_i v_j+\p_j v_i-\delta_{ij}\p_k v_k)^2 
	+\eta^b (\p_k v_k)^2 \,.
 \la{Q}
\eea
With this constitutive relations we easily derive using (\ref{energy-aux}) and (\ref{eq-rhou1}-\ref{eq-su1}}) the modified energy conservation as
\bea
	 \p_t\left[\epsilon +\frac{\rho v_k^2}{2}\right]
	&+& \p_i\left[(\epsilon+p)v_i +\frac{\rho v_k^2}{2}v_i
	-v_j\sigma_{ji}+J_i^E\right] 
	= \Omega \tau -\Gamma^v v_i^2-\Gamma^\Omega \Omega^2 \,.
\eea
The right hand side is the energy influx through the work by an external torque and the loss due to external friction forces. The correction to energy current is given by
\bea
	J_i^E &=& \left(\epsilon_\rho-\frac{v_k^2}{2}\right)J_i^\rho
	+T J_i^s+\Omega J_i^\ell \,.
\eea

The system of hydrodynamic equations (\ref{eq-rhou1}-\ref{eq-su1}) together with constitutive relations (\ref{Jirho}-\ref{Q}) gives a very general hydrodynamic description of system of active rotors. This description is characterized by many phenomenological constants and is too general for our purposes. In the following we will make simplifications to reduce the complexity of this system.

\section{Hydrodynamics of chiral active fluids with odd viscosity}
\label{sec:hyd}

In this paper we are looking for the effects related to the transport of angular momentum in the system of active rotors. We focus on the granular limit of hydrodynamics and neglect thermal effects assuming that the thermal physics does not play significant role. This means that we completely neglect all temperature dependences and omit (\ref{eq-su1}). We also choose a frame (a definition of velocity) such that $J_i^\rho=0$ and neglect all ``odd'' coefficients $L^{\mu,T,\Omega}=0$ except for $\Gamma$. This leaves us with much simpler hydrodynamic theory
\bea
	\p_t\rho+\p_i(\rho v_i) &=& 0\,,
 \la{eq-rhou2si} \\
	\p_t \ell+\p_i(\ell v_i) &=& \tau+D^\Omega \p_i^2\Omega
	- \Gamma^\Omega \Omega
	-\epsilon_{ij}\sigma_{ij} \,,
 \la{eq-lu2si} \\
	\p_t (\rho v_i) + \p_j(\rho v_i v_j ) &=& \p_{j}\sigma_{ij}-\Gamma^v v_i \,.
 \la{eq-rhouu2si} 
\eea
We see from (\ref{eq-lu2si}) that the anti-symmetric part of the stress $\sigma^{a}_{ij}=\epsilon_{ij}\Gamma(\Omega  - \omega)/ 2$ given by (\ref{sigmaa}) has a meaning of an internal torque acting between intrinsic rotational degrees of freedom and the rotational motion of the fluid. 
In Eq.~(\ref{eq-rhouu2si}), we have subsumed the pressure term into the definition of the fluid stress tensor.
For the case $\Gamma^v = 0$, Eqns.~(\ref{eq-rhou2si}-\ref{eq-rhouu2si}) are identical to Eqs.~(1-3) in the main text.

The complete equations that we simulate numerically, derived in the appropriate regimes in Secs.~\ref{sec:inc} and \ref{sec:com} are the equations (\ref{eq-rhou2si},\ref{eq-lu2si},\ref{eq-rhouu2si})
with
\begin{equation}
	\sigma_{ij} = - p \delta_{ij} + \eta \left(\partial_i v_j + \partial_j v_i-\delta_{ij}\p_{k}v_{k}\right)
	+ \eta^{odd}(\p_{i}v_{j}^{*}+\p_{i}^{*}v_{j})
	+ \frac{1}{2} \epsilon_{ij} \Gamma(\Omega  - \omega).
\end{equation}
that capture the effects of both odd viscosity $\eta^{odd}$ and antisymmetric stress $\Gamma$.

Equations (\ref{eq-rhou2si},\ref{eq-lu2si},\ref{eq-rhouu2si}) together with (\ref{sigmaa},\ref{sigmaijs}) (we will also put bulk viscosity $\eta^b=0$ and identify $\eta^{odd}=\ell/2$ in the following) is our starting point to study the effects of odd viscosity and anti-symmetric stress $\sigma^{a}$ on the dynamics of active rotors. 
The equations above are constructed phenomenologically. For a realistic system of active rotors one should either derive or at least estimate the values of various hydrodynamic parameters from the microscopic model. 

The angular momentum density is given by $\ell \equiv I \Omega$ with $I \equiv \iota \rho$ as the moment of inertia density ($\iota \sim a^2$, where $a$ is the linear size of the fluid's constituents particles). The active torque density $\tau$ is proportional to the density of rotors. It injects energy into the fluid, thus breaking detailed balance. The friction-coefficient $\Gamma^{\Omega}$ saturates energy injection so as to allow the fluid to reach a non-equilibrium steady state, whereas $D^\Omega$ controls diffusion of local rotations.

The crucial ingredient in Eqs.~(\ref{eq-rhou2si}-\ref{eq-rhouu2si}) is the anti-symmetric part of the stress $\sigma^{a}_{ij}=\epsilon_{ij}\Gamma \left(\Omega -\omega\right)/2$ that couples the flow and local rotation degrees of freedom. Here $\omega =\frac{1}{2}\p_k v_k^*$ is the local angular velocity of the rotation of the fluid. The coupling between $\Omega$ and $\vv$ through $\sigma^{a}$ terms of (\ref{eq-lu3},\ref{eq-rhouu3}) respects the corresponding Onsager relation and, therefore, does not inject additional energy into the system. These terms act only as frictional inter-rotor couplings that convert angular momentum between local rotation of the fluid particles and vorticity due to center-of-mass motion. This part is the same as in Ref.~\cite{Tsai2005}, but in addition to the terms in Ref.~\cite{Tsai2005} we also include the odd viscosity part of the stress tensor (\ref{sigmaijs}), proportional to the intrinsic angular momentum: $\eta^{odd}=\ell/2$.

We may also derive Eqs.~(\ref{eq-rhou2si}-\ref{eq-rhouu2si}) by introducing the functional $F[\mathbf{v},\Omega]$ as
\bea
	F[\mathbf{v},\Omega] = \int d\mathbf{x}\, \left\{
	\frac{\Gamma^v}{2}v_i^2+\frac{\Gamma^\Omega}{2}\Omega^2
	+\frac{D^\Omega}{2}(\p_i \Omega)^2\right\}\,.
 \la{F}
\eea
Using this non-negative functional we can rewrite (\ref{eq-lu2si},\ref{eq-rhouu2si}) as 
\bea
	\p_t \ell+\p_i(\ell v_i) &=& \tau-\sigma^{a}-\frac{\delta F}{\delta \Omega} \,,
 \la{eq-lu3} \\
	\p_t (\rho v_i) + \p_j(\rho v_i v_j -\sigma_{ij}^{s})  &=& \frac{1}{2}\epsilon_{ij}\p_{j}\sigma^{a}-\frac{\delta F}{\delta v_i}\,.
 \la{eq-rhouu3} 
\eea

\section{Incompressible hydrodynamics of chiral active fluids}
\label{sec:inc}
In this section, we consider the case in which the fluid is nearly incompressible. We first solve the equation of motion for the velocity field by using the incompressibility condition $\nabla \cdot \vv = 0$ and then substitute this result to find how the pressure $p$ deviates away from its steady-state value $p_0$. We then find the deviations in density by assuming small, linear compression, $\rho - \rho_0 = c^{-2} (p - p_0)$.
This approximation is valid as long as $c^{-2} |p - p_0| \ll \rho_0$.

Numerically, we keep the antisymmetric stress term to compare its effects with the effects of odd viscosity. Then, we solve the following system of equations for both $\vv$ and $\Omega$:
\begin{align}
\rho_0 \partial_t v_i + \rho_0 \nabla \cdot (v_i \vv) &= \eta \nabla^2 v_i + \frac{I}{2} \partial_j \left[ \Omega (\p_{i}v_{j}^{*}+\p_{i}^{*}v_{j}) \right] - \partial_i p + \frac{\Gamma}{2}\epsilon_{ij} \partial_j \left[ \Omega -\omega\right], \label{eq:numfig2}\\
I \partial_t \Omega + I \nabla \cdot (\Omega \vv)&= D^{\Omega} \nabla^2 \Omega - \Gamma^{\Omega} \Omega + \tau
 - \Gamma [\Omega - \omega]. \label{eq:om0}
\end{align}
We divide Eq.~(\ref{eq:numfig2}) by $\rho_0$ and Eq.~(\ref{eq:om0}) by $I$ and obtain
\begin{align}
\partial_t v_i + \nabla \cdot (v_i \vv) &= \nu \nabla^2 v_i - \rho_0^{-1} \partial_i p  + \frac{\iota}{2} \partial_j \left[ \Omega (\p_{i}v_{j}^{*}+\p_{i}^{*}v_{j}) \right]+ \frac{\Gamma^\prime}{2} \epsilon_{ij} \partial_j \left[ \Omega -\omega\right], \label{eq:numfig2-1}\\
\partial_t \Omega +\nabla \cdot (\Omega \vv)&= D^{\Omega\prime} \nabla^2 \Omega - \Gamma^{\Omega\prime} \Omega + \tau^{\prime}
 - \Gamma^{\prime} \iota^{-1} [\Omega - \omega]  \label{eq:om0-1}
\end{align}
where $\nu \equiv \eta/\rho_0$ is the kinematic viscosity, $\Gamma^\prime \equiv \Gamma/ \rho$, $D^{\Omega\prime} \equiv D^{\Omega}/I$, $\Gamma^{\Omega\prime} \equiv \Gamma^{\Omega}/ I$, $\tau^{\prime} \equiv \tau/I$, and $\iota \equiv I/ \rho_0$.

\subsection{Dimensionless parameters}
\label{sec:gas}

In the following, we will be solving Eqs.~(\ref{eq:numfig2-1}-\ref{eq:om0-1}) numerically. To understand better various regimes and to make further analytic progress it is useful to understand the dimensionless parameters governing the motion of the fluid.

Let us rescale all physical quantities to obtain a dimensionless problem.
From initial conditions of the form $\vv = v_0 \uv(t=0,\rv_r \equiv \rv/r_0)$, we obtain natural length and velocity scales. 
Then, we define dimensionless quantities (denoted by subscript $r$) via $\omega = \omega_r v_0/r_0$, $t =t_r r_0^2/\nu$, $p = p_r \nu v_0 \rho_0/r_0$,
$\Omega = \Omega_r \Omega_0$, where $\Omega_0 \equiv \tau/\Gamma^\Omega= \tau^{\prime}/\Gamma^{\Omega\prime}$ and find dimensionless equations,
{\footnotesize	
\begin{align}
	\partial_t {u}_i + \frac{v_0 r_0}{\nu}\nabla \cdot ({u}_i \uv) 
	&= \nabla^2 u_i -\partial_i p_r + 
	\frac{\iota \Omega_0}{2 \nu} \partial_j \left[ \Omega_r (\p_{i}u_{j}^{*}
	+\p_{i}^{*}u_{j}) \right]
	+ \frac{\Gamma^{\prime}}{2\nu}\frac{\Omega_0 r_0}{v_0}\epsilon_{ij} \partial_j \Omega_r
	- \frac{\Gamma^{\prime}}{2 \nu}\epsilon_{ij} \partial_j \omega_r, 
 \label{eq:numfig2-2}\\
	\partial_t  \Omega_r + \frac{v_0 r_0}{\nu}\nabla \cdot ( \Omega_r \uv)
	&= \frac{D^{\Omega\prime}}{\nu} \nabla^2  \Omega_r
	+ \frac{\Gamma^{\Omega\prime} r_0^2}{\nu} \left( 1 -  \Omega_r\right) 
	- \frac{\Gamma^{\prime}}{\nu}\frac{r_0^2}{\iota}  \Omega_r
	+ \frac{\Gamma^{\prime}}{\nu}\frac{r_0^2}{\iota}\frac{v_0}{\Omega_0 r_0} \omega_r,  \label{eq:om0-2}
\end{align} }
where all derivatives are now dimensionless, $v_0 r_0/\nu = \mathrm{Re}$ is the (small) Reynolds number, $\frac{\iota \Omega_0}{\nu} = 2 \nu_o/\nu$,
$\frac{\Gamma^{\prime}}{2\nu}\frac{\Omega_0 r_0}{v_0}$ determines the coupling of the velocity field to gradients in the $\Omega$ field due to the antisymmetric stress term,
$\frac{\Gamma^{\prime}}{2 \nu}$ determines the corrections to the dissipative viscosity due to the $\Omega$ field,
$\frac{D^{\Omega\prime}}{\nu}$ determines the relative importance of diffusivity for $\Omega$, $\frac{\Gamma^{\Omega\prime} r_0^2}{\nu}$
determines the strength of the friction for $\Omega$, $\frac{\Gamma^{\prime}}{\nu}\frac{r_0^2}{\iota}$ changes this friction via the antisymmetric stress,
$\frac{\Gamma^{\prime}}{\nu}\frac{r_0^2}{\iota}\frac{v_0}{\Omega_0 r_0}$ couples the vorticity to the $\Omega$ field.

\subsection{Integrating out the spinning frequency in the chiral gas regime}
\label{sec:gas2}
Although we solve Eqs.~(\ref{eq:numfig2-1}-\ref{eq:om0-1}) directly in our numerical computations, to make analytical progress, we consider the regime in which $\Omega \approx \mathrm{const}$. For this regime, the odd viscosity term dominates over the antisymmetric stress term. 

We assume that the angular velocity of intrinsic rotations is $\Omega_{0}\sim \tau/\Gamma^{\Omega}$ and that the parameter $\Gamma$ is sufficiently small so that there is a sufficiently large period of time over which one can consider particles rotating quickly and with almost no exchange between their intrinsic rotations and their center of mass motion. More precisely, we assume that $\omega\ll \Omega$ and comparing the odd viscosity term $\sim \eta^{odd} \nabla^{2} \vv^{*}$ and the antisymmetric stress terms $\sim \Gamma \nabla^{*}(\Omega-\omega)$ we require $\eta^{odd}\frac{v_0}{r_0} \gg \Gamma \Omega$ where $v_{0}$ and $r_{0}$ are typical velocity and length scales of the problem. Taking $\eta^{odd}\sim I\Omega$ we obtain the condition
\begin{equation}
 \label{eq:regime_0}
	\frac{\Gamma}{I} \ll \frac{ v_0}{r_0}\,.
\end{equation}
In addition, for Eq.~(\ref{eq:regime_0}) to be valid for all times, the antisymmetric stress term $\Gamma \omega$ in Eq.~(\ref{eq:om0-1}) must be smaller than the active torque $\tau$. 
Otherwise, the condition in Eq.~(\ref{eq:regime_0}) would be violated due to the coupling between intrinsic rotation and vorticity, and large spatial variations in $\Omega$ would occur. We require, therefore
\be
 \label{eq:regime_1}
	\frac{\tau}{\Gamma} \gg \frac{v_0}{r_0}\,.
\ee
The necessary condition for the existence of a regime with both Eqs.~(\ref{eq:regime_0},\ref{eq:regime_1}) satisfied is the following relation between hydrodynamics parameters:
\begin{equation}
 \label{eq:regime}
	\tau I \gg \Gamma^2\,.
\end{equation}

In this regime, we can assume that $\Omega \approx \Omega_0 = \mathrm{const}$. For this case and for (nearly incompressible) flows at low Mach number, Eq.~(\ref{eq:numfig2-1}) reduces to a modified Navier-Stokes equation with the addition of an odd viscosity term~\cite{Avron1998}:
\begin{equation}
	\partial_t v_i + (\mathbf{v}\cdot \bm\nabla)  v_i
	= \nu \nabla^2 v_i + \nu_o \nabla^2 \epsilon_{ij }v_j
	- \frac{1}{\rho_0} \partial_i p\,,
 \label{eq:lin3}
\end{equation}
i.e., Eq.~(5) of the main text. Note that in the incompressible case, the effects of odd viscosity can be subsumed by redefining the pressure via $p^{eff} \equiv p - 2 \eta^o \omega$.
We arrive at this conclusion by noting that $\nabla \cdot \vv = 0$ implies $ \nabla^2 \epsilon_{ij }v_j =  2 \p_{i} \omega$ and substituting this equation into Eq.~(\ref{eq:lin3}).
One can write
\begin{equation}
	\partial_t \mathbf{v} + (\mathbf{v}\cdot \bm\nabla)  \mathbf{v}
	= \nu \nabla^2 \mathbf{v} 
	-  \bm\nabla (p^{eff}/\rho_{0})\,,
 \label{eq:lin3p}
\end{equation}
which is identical in form to the conventional Navier-Stokes equation (see, e.g.,~\cite{Ganeshan2017}). One can find $p^{eff}$ from the flow and find the pressure of the fluid from $p = p^{eff} + 2 \eta^o \omega$.

Outside of the limit defined by Eqs.~(\ref{eq:regime_0},\ref{eq:regime_1}), there are contributions due to antisymmetric stress whose effects we examine numerically.
In the regime opposite to that set by Eqs.~(\ref{eq:regime_0}-\ref{eq:regime}), in which the antisymmetric stress dominates, the odd viscosity term is a small nonlinear correction to the hydrodynamics and can be neglected for the same reasons as it was neglected in Refs.~\cite{Tsai2005,Furthauer2012}.

\subsection{Analytic solution for the Lamb-Oseen vortex with odd viscosity}

Let us look for a radially symmetric vortex solution of Eq.~(\ref{eq:lin3}).
Taking the curl of Eq.~(\ref{eq:lin3}) we obtain the equations for vorticity $\omega = \frac{1}{2} \nabla \times \vv$:
\begin{equation}
	\partial_t \omega +(\mathbf{v}\cdot \bm\nabla)  \omega
	= \nu \nabla^2 \omega, 
 \label{eq:mfig2-om}
\end{equation}
This is a transport equation for vorticity with the diffusion-like term due to the shear viscosity. We consider an initial Gaussian vorticity profile, so that the azimuthal component of velocity is a function of the radius and the radial component is zero. Then, for initial conditions for the vorticity, we consider:
\begin{equation}
 	\omega(t = 0, r) = \frac{v_0}{r_0\pi} e^{-r^2/r_0^2}, 
 \label{eq:mfig2-om-ini}
\end{equation}
In this case, the full solution is given by
\begin{equation}
	\omega(t, r) = \frac{v_0 r_0}{4 \pi \nu (t_0 + t)} e^{-r^2/ 4 \nu (t_0 + t)}\,. 
 \label{eq:mfig2-om-ini2}
\end{equation}
where $t_0 \equiv r_0^2/(4 \nu)$.
From this expression for $\omega$, we obtain the velocity profile satisfying the relations $\frac{1}{2}\nabla \times \vv = \omega$ and $\nabla\cdot \vv = 0$ and find
\begin{equation}
	v_\theta(t, r) = \frac{v_0 r_0}{\pi r} \left[1 - e^{-r^2/ 4 \nu (t_0 + t)}\right]\,. 
 \label{eq:mfig2-om-ini3}
\end{equation}
This solution is identical to the conventional Lamb-Oseen solution as the odd viscosity does not enter Eq.~(\ref{eq:mfig2-om}). However, the resulting pressure is different.

We find the expression for the pressure from the radial equation of motion in polar coordinates:
\be
	\rho_0^{-1} \frac{\partial p^{eff}}{\partial r} = \frac{v^2_\theta}{r}.
\ee
Using $p^{eff} \equiv p - 2 \eta^o \omega$, we obtain
\be
	p- p_\infty = 2\eta^{o}\omega + \rho_{0}\int_{+\infty}^{r} dr^\prime\,\frac{v^2_\theta(r^\prime)}{r^\prime}\,,
\ee
where $p_\infty \equiv p(r = \infty)$.
Introducing $r_{s}(t) = r_{0}\sqrt{1 + t/t_{0}}$ and $p_{s}= \frac{\rho_{0}v_{0}^{2}}{\pi}$, we have
\be
	p - p_\infty= \frac{p_{s}}{1+t/t_{0}}\left[\frac{1}{2\pi}\int_{+\infty}^{r^2/r_{s}^2}dq\, \frac{(1-e^{-q})^{2}}{q^{2}}
	+\frac{2}{\mathrm{Re}^o}e^{-r^{2}/r_{s}^{2}}\right]\,,
\ee
where $\mathrm{Re}^o \equiv r_{0}v_{0}/\nu^{o}$ is an odd Reynolds number.

In the regime in which the Mach number $\mathrm{Ma} \equiv v_0/c \ll 1$, we use the relation $\rho - \rho_0 = c^{-2} (p - p_\infty)$ to calculate (small) changes in density as a result of the vortex flow:
\be
	\rho - \rho_0 = \frac{\rho_0 \mathrm{Ma}^2}{\pi^2(1+t/t_{0})}\left[\frac{1}{2}\int_{+\infty}^{r^2/r_{s}^2}dq\, \frac{(1-e^{-q})^{2}}{q^{2}}
	+\frac{2 \pi}{\mathrm{Re}^{o}}e^{-r^{2}/r_{s}^{2}}\right]\,.
\ee
Rescaling the density as 
\be
\delta \tilde{\rho} \equiv (\rho - \rho_0) \frac{\pi^2 (1+t/t_0)}{\mathrm{Ma}^2}
\ee
 and the radius as $\tilde{r} \equiv r/r_s$, we find that 
\be
	\frac{\delta \tilde{\rho}}{\rho_0} = \left[\frac{1}{2}\int_{+\infty}^{\tilde{r}^2}dq\, \frac{(1-e^{-q})^{2}}{q^{2}}
	+\frac{2 \pi}{\mathrm{Re}^{o}}e^{-\tilde{r}^2}\right]\,.
	\label{eq:rho-res}
\ee
We plot this solution (a rescaled form valid for all times $t$) in Fig.~2c of the main text.
In the case $\mathrm{Re}^o \ll 1$, we drop the first term of Eq.~(\ref{eq:rho-res}). Then, we 
plot $\delta \tilde{\rho}|\mathrm{Re}^{o}|/(2\pi \rho_0)  = \mathrm{\sign}( \nu^o)e^{-\tilde{r}^2}$ in the inset of Fig.~2c.
This is also the regime in which Eq.~(6) is valid.

Note that in addition to the condition $\mathrm{Ma} \ll 1$ that must be valid for these equations to hold, for small $\mathrm{Re}^o$ the self-consistency condition $\frac{|\rho - \rho_0|}{\rho_0} \ll 1$ also dictates that 
\begin{equation}
\frac{\mathrm{Ma}^2 }{|\mathrm{Re}^o|} \ll 1.
\end{equation}
Notably, an odd viscosity fluid can be compressible even at low Mach number if the 
odd Reynolds number $|\mathrm{Re}^o|$ is sufficiently small, or equivalently, if the odd viscosity is sufficiently large.

In Fig.~2b of the main text, we plot the rescaled solution for the azimuthal velocity, i.e., 
\be
\tilde{v}_\theta(\tilde{r})/v_0 = v_\theta(r) \pi r_s/r_0 = (1 - e^{-\tilde{r}^2})/\tilde{r}.
\ee

\section{Compression shocks in chiral active fluids}
\label{sec:com}
In this section, we are interested in the conditions for which Eq.~(\ref{eq-rhouu2si}) reduces to Burgers' equation. Under these conditions, the fluid experiences a compression shock,
but the density gradients are small. Schematically, we assume a lowest-order-nonlinearity and lowest-order-gradient expansion.
From this assumption, we note that terms of the form $\nabla \rho \nabla v$ are higher order in nonlinearity than the viscosity terms $\nabla^2 v$ 
and higher order in gradients than the nonlinear term $\nabla v^2$. Thus we may neglect these terms to derive a Burgers' equation with the addition of odd viscosity and antisymmetric stress.
In that case, the full equations of motion are:
\begin{align}
\partial_t \rho + \nabla \cdot (\rho {\bf v}) &= 0, \label{eq:rho3}\\
\partial_t (\rho v_i) + \nabla \cdot (\rho v_i \vv) &= \eta \nabla^2 v_i - \partial_i p + \frac{I}{2} \partial_j \left[ \Omega (\p_{i}v_{j}^{*}+\p_{i}^{*}v_{j}) \right]+ \frac{\Gamma}{2}\epsilon_{ij} \partial_j \left[ \Omega -\omega\right] + f_i(\rv), \label{eq:numfig3}\\
\iota \partial_t ( \rho \Omega) + \iota \nabla \cdot ( \rho \Omega \vv)&= D^{\Omega} \nabla^2 \Omega - \Gamma^{\Omega} \Omega + \tau
 - \Gamma [\Omega - \omega], \label{eq:om3}
\end{align}
where $f_i$ is the external forcing that gives rise to the steady-state shock.
We rewrite the left-hand side of Eqs.~(\ref{eq:numfig3},\ref{eq:om3}) using the product rule and the continuity Eq.~(\ref{eq:rho3}),
\begin{align}
\partial_t \rho + \nabla \cdot (\rho {\bf v}) &= 0, \label{eq:rho3-1}\\
\rho \left[ \partial_t v_i + (\vv\cdot\nabla) v_i \right] &= \eta \nabla^2 v_i - \partial_i p + \frac{I}{2} \partial_j \left[ \Omega (\p_{i}v_{j}^{*}+\p_{i}^{*}v_{j}) \right]+ \frac{\Gamma}{2}\epsilon_{ij} \partial_j \left[ \Omega -\omega\right] + f_i(\rv), \label{eq:numfig3-1}\\
\iota \rho \left[ \partial_t \Omega + (\vv\cdot\nabla) \Omega \right]&= D^{\Omega} \nabla^2 \Omega - \Gamma^{\Omega} \Omega + \tau
 - \Gamma [\Omega - \omega] \label{eq:om3-1}
\end{align}
We divide both sides of Eq.~(\ref{eq:numfig3-1}) by $\rho$ and both sides of Eq.~(\ref{eq:om3-1}) by $\rho \iota$ and Taylor-expand the (dynamic) hydrodynamics coefficients in density, keeping only the lowest-order (constant) terms to find a set of equations similar to Eqs.~(\ref{eq:numfig2-1},\ref{eq:om0-1}):
\begin{align}
\partial_t \rho + \nabla \cdot (\rho {\bf v}) &= 0, \label{eq:rho3-2} \\
\partial_t v_i + \nabla \cdot (v_i \vv) &= \nu \nabla^2 v_i - \rho^{-1}\partial_i p + \frac{\iota}{2} \partial_j \left[ \Omega (\p_{i}v_{j}^{*}+\p_{i}^{*}v_{j}) \right]+ \frac{\Gamma^\prime}{2} \epsilon_{ij} \partial_j \left[ \Omega -\omega\right] +  f^\prime_i(\rv), \label{eq:numfig3-2}\\
\partial_t \Omega +\nabla \cdot (\Omega \vv)&= D^{\Omega\prime} \nabla^2 \Omega - \Gamma^{\Omega\prime} \Omega + \tau^{\prime}
 - \Gamma^{\prime} \iota^{-1} [\Omega - \omega], \label{eq:om3-2}
\end{align}
where $f^\prime_i = f_i/\rho$. Eqs.~(\ref{eq:rho3-2}-\ref{eq:om3-2}), along with the equation of state $p(\rho)$, describe the full hydrodynamics.
In the limit of strong compression, we may drop the pressure term and Eqs.~(\ref{eq:numfig3-2},\ref{eq:om3-2}) then form a closed set,
\begin{align}
	\partial_t v_i + \nabla \cdot (v_i \vv) &= \nu \nabla^2 v_i 
	+ \frac{\iota}{2} \partial_j \left[ \Omega (\p_{i}v_{j}^{*}+\p_{i}^{*}v_{j}) \right]
	+ \frac{\Gamma^\prime}{2} \epsilon_{ij} \partial_j \left[ \Omega -\omega\right] 
	+  f^\prime_i(\rv), 
 \label{eq:numfig3-3}\\
	\partial_t \Omega +\nabla \cdot (\Omega \vv)
	&= D^{\Omega\prime} \nabla^2 \Omega - \Gamma^{\Omega\prime} \Omega 
	+ \tau^{\prime} - \Gamma^{\prime} \iota^{-1} [\Omega - \omega] \,.
 \label{eq:om3-3}
\end{align}

After solving Eqs.~(\ref{eq:numfig3-3},\ref{eq:om3-3}), the density profile can be found using the continuity equation Eq.~(\ref{eq:rho3-2}). 
We rescale these equations to find the dimensionless form. In the steady state, the characteristic length and velocity scales come
from the forcing term $f^\prime_i(\rv) = (v_0^2/r_0) f^\prime_r (\rv/r_0)$. We also rescale $t =t_r r_0^2/\nu$,
$\Omega = \Omega_r \Omega_0$ the same way as for Eqs.~(\ref{eq:numfig2-2},\ref{eq:om0-2}). We find a set of equations that describe the strongly nonlinear regime:
{\footnotesize	
\begin{align}
\partial_t {u}_i + \frac{v_0 r_0}{\nu}\nabla \cdot ({u}_i \uv) 
	&= \nabla^2 u_i + \frac{v_0 r_0}{\nu} f_r^\prime (\rv_r) 
	+ \frac{\iota \Omega_0}{2 \nu} \partial_j \left[ \Omega_r (\p_{i}u_{j}^{*}
	+\p_{i}^{*}u_{j}) \right]
	+ \frac{\Gamma^{\prime}}{2\nu}\frac{\Omega_0 r_0}{v_0}\epsilon_{ij} \partial_j \Omega_r
	- \frac{\Gamma^{\prime}}{2 \nu}\epsilon_{ij} \partial_j \omega_r,
 \label{eq:numfig3-4}\\
	\partial_t  \Omega_r + \frac{v_0 r_0}{\nu}\nabla \cdot ( \Omega_r \uv)
	& = \frac{D^{\Omega\prime}}{\nu} \nabla^2  \Omega_r
	+ \frac{\Gamma^{\Omega\prime} r_0^2}{\nu} \left( 1 -  \Omega_r\right) 
	- \frac{\Gamma^{\prime}}{\nu}\frac{r_0^2}{\iota}  \Omega_r
	+ \frac{\Gamma^{\prime}}{\nu}\frac{r_0^2}{\iota}\frac{v_0}{\Omega_0 r_0} \omega_r.
 \label{eq:om3-4}
\end{align}}
Note that Eq.~(\ref{eq:om3-4}) is identical to Eq.~(\ref{eq:om0-2}), whereas Eq.~(\ref{eq:numfig3-4}) is Eq.~(\ref{eq:numfig2-2}), but with the forcing term instead of the pressure term. The other crucial feature of Eqs.~(\ref{eq:numfig3-4},\ref{eq:om3-4}) is that the incompressibility condition $\nabla \cdot \uv = 0$ does not apply, unlike for Eqs.~(\ref{eq:numfig2-2},\ref{eq:om0-2}).

Given a forcing ${{\bf f}}_r^\prime (\rv_r) = (f^\prime_{rx} (\rv_r), 0 )$, we find a steady state that depends only on the $x$-coordinate. This allows us to solve for the density profile using the relation $\partial_x \rho/\rho = - \partial_x v_x/v_x$ [from Eq.~(\ref{eq:rho3-2})], which along with the mean value $\rho_0$ of the density, allows us to calculate the profile $\rho(\xv)$ using either the numerical (Fig.~3a) or analytical (Figs.~3b-f) solutions for $v_x(x)$.

\subsection{Shocks: integrating out the spinning frequency in the chiral gas regime}
We now follow the same logic as was used in Sec.~\ref{sec:gas2}, but without assuming incompressibility. We use the same conditions on initial scales (\ref{eq:regime_0},\ref{eq:regime_1}). This regime is only possible when the condition on the hydrodynamic parameters (\ref{eq:regime}) is satisfied. In this case we can assume that $\Omega \approx \Omega_0 = \mathrm{const}$.

As in Sec.~\ref{sec:gas2}, outside of the limit defined by Eqs.~(\ref{eq:regime_0},\ref{eq:regime_1}), there are contributions due to antisymmetric stress whose effects we examine numerically.
In the regime opposite to that set by Eqs.~(\ref{eq:regime_0},\ref{eq:regime_1}), in which the antisymmetric stress dominates, the odd viscosity term is a small nonlinear correction to the hydrodynamics and can be neglected for the same reasons as it was neglected in Refs.~\cite{Tsai2005,Furthauer2012}.

In the case for which odd viscosity dominates, Eq.~(\ref{eq:numfig3-3}) reduces to a modified Burgers' equation with the additional odd viscosity term:
\begin{align}
	\partial_t v_i + (\vv \cdot \nabla) v_i 
	&= \nu \nabla^2 v_i + \nu_o \nabla^2 \epsilon_{ij} v_j
	+ f^\prime_i (\rv) \,. 
 \label{eq:numfig3-5}
\end{align}

Eq.~(\ref{eq:numfig3-5}) may be re-written as an equation for the evolution of vorticity, in which $\nu^o$ contributes an additional source term as
a result of the compressible part $\nabla \cdot \vv$ of the flow (provided that $\nabla \times \mathbf{f}=0$):
\begin{equation}
	\partial_t \omega + \nabla \cdot (\omega \vv) = \nu \nabla^2 \omega + \frac{\nu_o}{2 } \nabla^2 (\nabla \cdot \vv)\,. 
 \label{eq:numfig3-6si}
\end{equation}
As explained in the main text, the odd viscosity generates vorticity preferentially within the shock where gradients of $\nabla \cdot \vv$ are largest.
The dimensionless version of Eq.~(\ref{eq:numfig3-5}) reads ($\mathrm{Re}=\frac{v_0 r_0}{\nu}$)
\begin{align}
	\partial_t u_i + \mathrm{Re}\, (\uv \cdot \nabla) u_i 
	&= \nabla^2 u_i + \frac{\nu_o}{\nu} \nabla^2 \epsilon_{ij} u_j
	+ \mathrm{Re}\, {f}^\prime_{ri} ({\rv}_r) \,. 
 \label{eq:numfig3-7}
\end{align}

In the steady state, the one-dimensional profile for Eq.~(\ref{eq:numfig3-7}) is determined by
\begin{align}
	\mathrm{Re}\, u_x \partial_x u_x 
	&= \partial_x^2 u_x + \frac{\nu_o}{\nu} \partial_x^2 u_y
	+ \mathrm{Re}\, f^\prime_{rx} (x_r) \,, 
 \label{eq:vx1}\\
	\mathrm{Re}\, u_x \partial_xu_y 
	&= \partial_x^2 u_y - \frac{\nu_o}{\nu} \partial_x^2 u_x \,. 
 \label{eq:vy1}
\end{align}

\subsection{Perturbative solution for small odd viscosity}{

For the case $\nu_o/\nu \ll 1$, we obtain a perturbative analytical solution for the steady-state shock profile by first neglecting the subdominant $\nu_o$ term in Eq.~(\ref{eq:vx1}),
solving this equation both within and outside the shock using matched assymptotics, and then substituting the solution inside the shock into Eq.~(\ref{eq:vy1}).
We choose the forcing to be ${{\bf f}}_r^\prime ({\rv}_r) = (- \pi \sin ({x}_r \pi ) /4, 0 )$
and solve the equation looking for the periodic solution with the period $x_r\in [-1,1]$. We solve
\begin{equation}
	u_x \partial_x u_x =  \nu_r \partial_x^2 u_x 
	- \frac{\pi}{4} \sin( x_r \pi ) \,, 
 \label{eq:vx2}
\end{equation}
where $\nu_r = \nu/(v_0 r_0)$ is the corresponding inverse Reynolds number.
For large Reynolds number ($\nu_r\ll 1$), the steady-state solution has a narrow shock in the vicinity of ${x}_r = 0$.
Away from this region, the inertial term dominates and the steady-state velocity profile is obtained by integrating the 
expression $\partial_x u_x^2 = - \frac{\pi}{2} \sin( x_r \pi )$ to find the solution
$u_x = - \mathrm{sign}(x_r)\cos( x_r \pi /2 )$.  Here we tuned the integration constant so that the shock is at $x_r=0$ and the velocity is zero on average.

In the region $x_r/\nu_r \ll 1$, a different solution applies. 
In that region, Eq.~(\ref{eq:vx2}) can be simplified by dropping the forcing term, and then integrated exactly using matched-asymptotic boundary conditions $u_x \rightarrow 1$ for $x_r/\nu_r \ll -1$, and $u_x \rightarrow - 1$ for $x_r/\nu_r \gg 1$. The resulting solution is given by $u_x = -\tanh [x_r/ (2 \nu_r)]$.
For the case $\nu_r \ll 1$ a simple interpolation, 
\begin{equation}
	u_x (x) = -\tanh [x_r/ (2 \nu_r)]\cos( x_r \pi /2 )
\end{equation}
between the outer and the inner solution gives a reasonable approximation to the exact steady state over the entire range of values of $x_r$.

The solution for $u_y$ decays rapidly away from the shock, so to find the steady-state $u_y$ profile,
it is sufficient to consider the inner solution for $u_x$.
We substitute this solution into ($\nu_r^{o} = \nu^{o}/(v_0 r_0)$)
\begin{equation}
	u_x \partial_x u_y 
	= \nu_r \partial_x^2 u_y - \nu_r^o \partial_x^2 u_x \,, 
 \label{eq:vy2}
\end{equation}
and find a \emph{linear} ODE for $u_y$:
\begin{equation}
	\nu_r \partial_x^2 u_y + \tanh \left(\frac{x_r}{2 \nu_r}\right) \partial_xu_y 
	= \frac{\nu_r^{o}}{2 \nu_r^2} \sech^2 \left(\frac{x_r}{2 \nu_r}\right)
	\tanh\left(\frac{x_r}{2 \nu_r}\right). 
 \label{eq:vy30}
\end{equation}
We multiply the above equation by $\cosh^2\left(\frac{x_r}{2 \nu_r}\right)$ and integrate, which leads to:
\begin{equation}
	\frac{d u_y}{d x_r} = \frac{\nu_r^{o}}{\nu_r^2} \sech^2\left(\frac{x_r}{2\nu_r}\right) 
	\ln \left| e^{-c_{1}}\cosh \left(\frac{x_r}{2\nu_r}\right)\right| \,,
\end{equation}
where  $c_1$ is an arbitrary constant. We then integrate this expression again, to find
\begin{equation}
	u_y(x_r)  = - \frac{{ 2} \nu_r^{o}}{\nu_r} \left[\frac{x_r}{2\nu_r} 
	- \tanh\left(\frac{x_r}{2 \nu_r} \right) 
	\left( \ln \left| e^{1-c_{1}}\cosh \left(\frac{x_r}{2\nu_r}\right)\right|  \right) \right] + c_2 \,.
\end{equation}
The boundary conditions $u_y \rightarrow 0$ as $x/(2\tilde{\nu}) \rightarrow \pm \infty$ require the choice of integration constants $c_1 =1 - \ln 2$ and $c_2 = 0$.
Substituting these values and simplifying, we find:
\begin{equation}
	u_y(x_r) 
	= - \frac{{ 2} \nu_r^{o}}{\nu_r} \left[\frac{x_r}{2\nu_r} 
	- \tanh\left(\frac{x_r}{2 \nu_r} \right) 
	\left( \ln \left| 2\cosh \left(\frac{x_r}{2\nu_r}\right)\right| \right) 
	\right].
\end{equation}
We then re-express this solution in terms of the viscosity ratio $\nu_o / \nu$ and Reynolds number $\mathrm{Re}$.
\begin{equation}
	\frac{\nu}{2 \nu_{o}}u_y(x_r) = -\hat{x}+\tanh \hat{x}\,\left(\ln\Big|2\cosh\hat{x}\Big|\right)\,,
\end{equation}
where $\hat{x}=x\mathrm{Re}/2$.
We plot this solution in Fig.~3a alongside the numerical solution to the full Eqs.~(\ref{eq:numfig3-5}) once the system reaches the steady state.

\subsection{Scaling for large odd viscosity}

For the case $|\nu_o|/\nu \gg 1$, we use scaling to obtain the characteristic features of the profile.
The equations that determine the steady-state profile are:
\begin{align}
	u_x \partial_x u_x &= \nu_r \partial_x^2 u_x + \nu_r^{o} \partial_x^2 u_y - \frac{\pi}{4} \sin( x_r \pi ) \,,
 \label{eq:vx3} \\
	u_x \partial_x u_y &= \nu_r \partial_x^2 u_y - \nu_r^{o} \partial_x^2 u_x \,. 
 \label{eq:vy3}
\end{align}
Let us drop the forcing and $\nu$ terms. We obtain
\begin{align}
	u_x \partial_x u_x &=  \nu_r^{o} \partial_x^2 u_y  \,,
 \label{eq:vx30} \\
	u_x \partial_x u_y &= - \nu_r^{o} \partial_x^2 u_x \,. 
 \label{eq:vy31}
\end{align}
Integrating the first equation gives $2 \nu^o_r \p_{x}u_y = u_x^{2}-C_{1}$, where $C_{1} = 1$ for the case we consider. Substituting into the second equation we obtain
$$
	u_x(u_x^{2}-1) = -2 (\nu_r^o)^2 \p_{x}^{2}u_x\,.
$$
This equation describes the motion of a nonlinear pendulum. We now linearize in $\delta u = u_x - 1$ to find the harmonic oscillator equation
\be
 \delta u = - (\nu^o_r)^2 \p_x^2 (\delta u).
\ee
The solutions to this equations oscillate in space with a period given by $\lambda \sim \nu^o_r$, which is the scaling law that we observe numerically in Fig.~3c.

If we take Eqs.~(\ref{eq:vx3}-\ref{eq:vy3}) and drop the forcing term, but not the $\nu$ terms, performing the same operations of integration we obtain the equation
\be
 \nu_r^o \p_x u_y = \frac{1}{2} (u_x^2 - 1) - \nu_r \p_x u_x.
\ee 
Substituting for gradients of $u_y$ in Eq.~(\ref{eq:vy3}), we then obtain
\be
 	\frac{1}{2} u_x (u_x^2 - 1) + [(\nu_r^o)^2 + \nu_r^2] \p_x^2 u_x = \nu_r \p_x u_x^2.
 \la{eq:nlin1}
\ee
Note that this is a nonlinear damped oscillator, with the damping term on the right-hand side.
Linearizing in $\delta u = u_x - 1$, we find
\be
 [(\nu_r^o)^2 + \nu_r^2] \p_x^2 (\delta u) - 2 \nu_r \p_x \delta u + \delta u = 0,
\ee
which describe the motion of a damped harmonic oscillator. From this, we find the characteristic wavelength to be (from comparing first and last terms) 
\be
 \lambda^2 \sim [(\nu^o)^2 + \nu^2]/v_0^2
\ee
and the decay length of the envelope from the damping ratio (from comparing first and second terms):
\be
 \Lambda \sim [(\nu^o)^2 + \nu^2]/\nu v_0.
\ee
In the limit $|\nu_o|/\nu \gg 1$, the oscillation wavelength is given by $\lambda \sim |\nu^o|/v_0$ (same as above) and the envelope size is $\Lambda \sim (\nu^o)^2/\nu v_0 \sim |\nu^o| \lambda/ \nu$. We find that these scaling laws are consistent with the full numerical solution of Eqs.~(\ref{eq:numfig3-5}), see inset of Fig.~3c. As expected, the envelope size diverges as viscosity goes to zero. In the opposite limit, $|\nu_o|/\nu \ll 1$, also considered in the previous section, we recover the same scaling laws: $\lambda \sim \Lambda \sim \nu /v_0$. In that limit, the oscillator is critically damped and no oscillations are observed.

\section{Numerical simulations}

In this section, we discuss the numerical method used for the direct numerical simulations of
hydrodynamic equations. The equations are solved using the pseudo-spectral method.
This method involves calculating spatial derivatives
in Fourier space and non-linear terms in real space.  The spatial
derivatives are non-local in real space and local in Fourier space, while the
non-linearities are local in real space but non-local in Fourier space. Thus, in
this method we restrict ourselves in evaluating local terms in both real and
Fourier space. The time integration is done to
the (spatially) Fourier-transformed fields. To calculate these Fourier transforms
we use the open-source library fftw-2.1.5.

In order to understand the implementation of the algorithm let us look at the
various kind of terms that need to be implemented. The Laplacian operator in Fourier space is given by $\nabla^2 \rightarrow -k^2$.  Let us now consider the (advection like) non-linear terms
of the form $\nabla \cdot (\rho \vv)$.  The product $\rho \vv$
is a convolution in Fourier space and highly non-local, but in real space this
is a local product. Therefore, the product is calculated in real space and the
divergence of the term evaluates to $i k_x \widehat{{\rho}v_x} +
i k_y\widehat{{\rho}v_y}$. 

Now let us consider an equation of the form $\partial_t \rho + \nabla
\cdot (\rho \vv) = D^{\rho} \nabla^2 \rho$.  In order to solve this
pseudo-spectrally, we take the spatial Fourier transform of the equation and we
are left with $\partial_t {\hat \rho} +i k_x \widehat{{\rho}v_x} + i k_y
\widehat{{\rho}v_y} = -D^{\rho} k^2 {\hat \rho}$.  It is possible to further
simplify the equation by rearranging the terms and calculating the integration
factor.  We are finally left with: $\partial_t e^{D^{\rho} k^2 t} {\hat \rho} =
-e^{D^{\rho} k^2 t} \left(i k_x \widehat{{\rho}v_x} + i k_y \widehat{{\rho}v_y}
\right)$. The solution of the above equation is :
\begin{equation}
{\hat \rho}(t) = -e^{-D^{\rho} k^2 t} \int_{t_0}^t dt' e^{D^{\rho} k^2 t'} i {\bf k} \cdot \widehat{{\rho}{\vv}}.
\end{equation}

The time evolution is done using a second order Runge-Kutta (RK2) method. The
pseudo-spectral method has the additional advantage of having exponentially
small spatial truncation error as opposed to algebraically small error which would result were we to use
a finite-difference scheme. The time evolution error in RK2 scheme is of
the order of $\sim \mathcal{O}(\delta t)^2$. 

One of the key concerns on calculating the convolutions numerically is the
generation of aliasing errors. In our simulations we cut off wave numbers 
greater than $2 K/3$ where $K\equiv N/2$ corresponds to the smallest lengthscales in our system ($N$ is the number of grid points).

\subsection{Incompressible regime}

We solve the incompressible Navier-Stokes equations coupled to
the local rotation field. We define a new tensor $\phi_{ij} = \partial_i v_j^{*} + \partial_i^{*} v_j$.
The incompressible Navier Stokes
can be further simplified using the constraint $\nabla \cdot \vv = 0$. Taking the curl of Eq.~(\ref{eq:numfig2-1}). The
resulting equations are:
\begin{eqnarray}
\partial_t \omega + \partial_l \left( v_l \omega \right) &=& \nu \nabla^2 \omega + \frac{\iota}{2} \epsilon_{ik} \partial_k \partial_j \left( \Omega \phi_{ij} \right)
+ \frac{\Gamma^{\prime}}{2} \nabla^2 [\Omega -\omega]. \\
\partial_t \Omega +  \partial_l ( v_l \Omega) &=& D^{\Omega \prime} \nabla^2 \Omega - \Gamma^{\Omega \prime} \Omega + \tau^{\prime} - \Gamma^{\prime} \iota^{-1} (\Omega -\omega).
\end{eqnarray}

The above equations can be written in the form discussed above in the Fourier space to be integrated
over time numerically. The form of the solutions are:
\begin{eqnarray}
{\hat \omega} (t) &=& -e^{-\nu k^2 t} \int_{t_0}^t dt' e^{\nu k^2 t'} [i k_l \widehat{v_l{\omega}} - \frac{\Gamma^{\prime}}{2} k^2 ({\hat \Omega} - {\hat \omega}) - \frac{\iota}{2} \epsilon_{ij} k_k k_j \widehat{\Omega \phi_{ij}}], \\
{\hat \Omega} (t) &=& -e^{-(D^{\Omega \prime} k^2 + \Gamma^{\Omega \prime})  t} \int_{t_0}^t dt' e^{(D^{\Omega \prime} k^2 + \Gamma^{\Omega \prime}) t'}  [i k_l \widehat{v_l {\Omega}} + \frac{\Gamma^{\prime}}{\iota} ({\hat \Omega} - {\hat \omega})].
\end{eqnarray}

Now, pressure can be calculated by taking the divergence of Eq.~(\ref{eq:numfig2-1})
\begin{equation}
\rho_0^{-1} \nabla^2 p^{eff} = - \partial_i [\nabla \cdot (\vv v_i)] + \frac{\iota}{2} \partial_i \partial_j \left( \Omega \phi_{ij} \right).
\end{equation}

The above equations provide us the information presented in the Figs.~2b and 2d in the main paper. The dimensionless parameters of the
simulations are
 $\left(\frac{v_0 r_0}{\nu}, \frac{\nu_o}{\nu}, \frac{\Gamma^{\prime}}{2\nu}, \frac{\Omega_0 r_0}{v_0},
\frac{D^{\Omega\prime}}{\nu},\frac{\Gamma^{\Omega\prime} r_0^2}{\nu},\frac{r_0^2}{\iota}\right) = (0.05,\pm 0.01,0.25, 0.004,1,0.1,1)$ on a grid of size $L/ r_0 = 20 \pi$ and lattice spacing $a/r_0 = 0.2$ over a time $\tilde{t} = 1.0$.  As initial conditions we use: $\tilde{\omega} = \frac{e^{-\tilde{r}^2}}{\pi}$ and $\tilde{\Omega} = 0$.

\subsection{Highly compressible regime}

The two-dimensional Burgers' equation [Eqs.~(\ref{eq:numfig3},\ref{eq:om3})] that is used to model compressible flow is similarly solved
by the pseudo-spectral algorithm. The data for Fig.~3 is obtained by solving this equation. The integral form of the solutions in Fourier space is :
\begin{equation}
{\hat v_i} (t) = -e^{-\nu k^2 t} \int_{t_0}^t dt' e^{\nu k^2 t'} [i {\bf k} \cdot \widehat{{\vv} v_i} + \nu_o k^2 \epsilon_{ij} {\hat v_j} 
+ i \frac{\iota}{2} k_j \widehat{{\Omega} \phi_{ij}} 
+ i  \frac{\Gamma^{\prime}}{2} \epsilon_{ij} k_j ({\hat \Omega} - {\hat \omega})].
\end{equation}
The solutions for $\Omega$ is the same as that for the incompressible flow. 

We simulate Eqs.~(\ref{eq:numfig3-3},\ref{eq:om3-3}) with an applied forcing $f^\prime_r(\rv)=\sin(x)/2$ for $x$ is in $[0,2\pi]$ (we rescale all lengths according to $x \rightarrow (x - \pi)/\pi$ in our results). We use parameters $\left(v_0 r_0/\nu, \nu_o/\nu, \frac{\Gamma^{\prime}}{2\nu}, \frac{\Omega_0 r_0}{v_0},
\frac{D^{\Omega\prime}}{\nu},\frac{\Gamma^{\Omega\prime} r_0^2}{\nu},\frac{r_0^2}{\iota}\right) = (44, -0.02, 2.0, 2.2, 2.0, 1800, 9)$ with lattice spacing $a = 0.006$.
For the simulations with high odd viscosity ratio (Fig.~3c-d of the main paper) we use $\nu_o/\nu = -10$.

\end{document}